\documentclass[prl,twocolumn,10pt]{revtex4-1}
\usepackage[pagebackref=false,pdfpagemode=UseNone,bookmarksopen=false,colorlinks=true,linkcolor=blue,citecolor=blue]{hyperref}
\usepackage{graphicx, bm, amsfonts, amssymb, amsmath, amsthm, color}
\usepackage[T1]{fontenc} 

\newcommand{\Dens}{\mathcal{D}} 
\newcommand{\Diag}{\Lambda} 
\newcommand{\Vari}{\mathcal{V}} 
\newcommand{\x}{O}
\newcommand{\Sre}{S} 
\newcommand{\Dkl}{D} 

\newcommand{\Q}{\mathcal{Q}}
\newcommand{\X}{\mathcal{X}}
\newcommand{\II}{\mathbb{I}}
\newcommand{\RR}{\mathbb{R}}
\DeclareMathOperator*{\tr}{\mathrm{tr}}
\DeclareMathOperator*{\ran}{\mathrm{ran}}
\DeclareMathOperator*{\arginf}{\arg\inf}
\newcommand*{\arginff}[2]{\arginf_{#1}\left\{#2\right\}}
\newcommand{\ra}{\rightarrow}
\newcommand{\ii}{\mathrm{i}}
\newcommand{\dd}{\mathrm{d}}
\newcommand{\ee}{\mathrm{e}}
\newcommand*{\inlinequotation}[1]{\guillemotleft{#1}\guillemotright}
\begin{document}
\title{Quantum collapse rules from the maximum relative entropy principle}

\author{Frank Hellmann}
 \email[Electronic Address: ]{Hellmann@pik-potsdam.de}
 \affiliation{Albert Einstein Institute, Golm, Germany}
 \affiliation{Potsdam Institute for Climate Impact Research, Potsdam, Germany}

\author{Wojciech Kami\'{n}ski}
 \email[Electronic Address: ]{Wojciech.Kaminski@fuw.edu.pl}
 \affiliation{Institute for Theoretical Physics, Faculty of Physics, University of Warsaw, Warszawa, Poland}

\author{Ryszard Pawe{\l} Kostecki}
 \email[Electronic Address (corresponding author): ]{Ryszard.Kostecki@fuw.edu.pl}
 \affiliation{Perimeter Institute for Theoretical Physics, Waterloo, Canada\\\ }

\date{August 6, 2014}

\begin{abstract}
\noindent We show that the von Neumann--L\"{u}ders collapse rules in quantum mechanics always select the unique state that maximises the quantum relative entropy with respect to the premeasurement state, subject to the constraint that the postmeasurement state has to be compatible with the knowledge gained in the measurement. This way we provide an information theoretic characterisation of quantum collapse rules by means of the maximum relative entropy principle.

\vskip+2mm\noindent{\scriptsize \textbf{keywords:} quantum measurement, maximum entropy, quantum information, quantum collapse.\\
\textbf{published:} New J. Phys. \textbf{18}, 013022 (2016).}
\end{abstract}

\maketitle

\section{Introduction}
%

The dynamics of quantum states in the orthodox (von Neumann's) foundations of quantum mechanics consist of two different prescriptions: the unitary evolution and the so-called `collapse' of a quantum state to a subspace encoding the knowledge gained in the outcome of a measurement. The mappings (rules) describing this collapse were originally formulated by von Neumann \cite{vonNeumann:1932:grundlagen} and later improved by L\"{u}ders \cite{Lueders:1951}. There are two different forms of collapse. When one knows only that a measurement corresponding to an observable (a self-adjoint operator with a discrete spectrum) $\x$ has taken place, the `weak' rule applies. It is defined as $\rho\mapsto\sum_{i\in I}P_i\rho P_i$, where $\rho$ is the original quantum state (in general, a density operator), while $\x=\sum_{i\in I}\lambda_iP_i$ is a spectral decomposition with some countable index set $I$ (hence, $\sum_{i\in I}P_i=\II$, $P_iP_j=P_i\delta_{ij}$, and $\lambda_i\in\RR$ $\forall i,j\in I$). If a measurement corresponding to $\x$ has resulted in a specific value $\lambda_k\in\{\lambda_i\mid i\in I\}$ associated to a projector $P_k\in\{P_i\mid i\in I\}$, then the `strong' rule, $\rho\mapsto P_k\rho P_k/\tr(\rho P_k)$, is applied.

The negative of Umegaki's quantum relative entropy \cite{Umegaki:1961,Umegaki:1962}, $\Dkl(\rho,\sigma) = - \Sre(\rho,\sigma) :=\tr(\rho\ln\rho-\rho\ln\sigma)\in[0,\infty]$, can be used as a measure of distinguishability, or relative information content, of the quantum state $\sigma$ from the state $\rho$. The use of $\Dkl$ instead of $\Sre$ follows Wiener's idea that the \inlinequotation{amount of information is the negative of the quantity defined as entropy} \cite{Wiener:1948}. Note that we call $\Sre = -\Dkl$ the relative entropy, following the convention of \cite{Bratteli:Robinson:1979} that makes the Gibbs--Shannon and von Neumann entropies the special cases of $\Sre$, after adding a constant: $\Sre_{\mathrm{vN}}(\rho)=\Sre(\rho,\II/n)+\log(n)$.

The function $\Dkl$ can be considered as a nonsymmetric distance: in general, $\Dkl(\rho,\sigma)\neq \Dkl(\sigma,\rho)$. If a given state is $\sigma$ and we believe it to be $\rho$, it can be easier or harder to find our error than if their roles were reversed. Say, $\sigma = P$ with $P$ some projector and $\rho=\II/n$. If we measure the property corresponding to $\II-P$, a single measurement can tell us that the state is not $\sigma$, whereas no single measurement could reveal the same of $\rho$. See e.g. \cite{Schumacher:Westmoreland:2002,Vedral:2002} for an overview of reasons for using $\Dkl(\rho,\sigma)$ as a measure of distinguishability and relative information content.

A key information theoretic property of the strong collapse rule is that the probability of measuring the value $\lambda_k$ again, after having measured it once, is $1$, which follows from $\tr\left(P_k\frac{P_k\rho P_k}{\tr(\rho P_k)}\right)=1$. Repeated measurements add no new information. Clearly, the state $P_k\rho P_k/\tr(\rho P_k)$ is not the only state that has this property (note that $P_k$ is not necessary a rank 1 projector). What we demonstrate in this letter is that, among all states that have this property, the strong collapse rule selects the state that is least distinguishable from the initial state $\rho$, that is, it has the minimum relative information $\Dkl(\rho,\cdot)$, in a suitably regularised sense. This allows for an information theoretic characterisation of the strong collapse rule: the state after measurement is the state that is least distinguishable from the previous state, while being compatible with the new information gained by the measurement.

In order to derive the strong collapse rule, we will need two intermediate results. First we will show that the weak collapse rule produces the least distinguishable state among the block diagonal states. We then show that a weighted version of the strong collapse rule, $\rho\mapsto\sum_ip_i P_i\rho P_i/\tr(\rho P_i)$, is the least distinguishable amongst the states with blocks of fixed trace. This rule can be interpreted as corresponding to a measurement where we believe that the result $P_i$ occurred with probability $p_i$. This intermediate step regularises the problem of a strong collapse, which is then obtained as a limiting case, by taking $p_i \mapsto \delta_{ik}$ with $k=1$.

Our derivation of the collapse rules from the constrained maximisation of Umegaki's quantum relative entropy is of special importance in the context of epistemic and information theoretic approaches to the foundations of quantum theory. In this context, collapse rules have been considered as analogues of the Bayes--Laplace rule \cite{Bub:1977,Bub:1979:measurement,Fuchs:2002,Jacobs:2002}. This analogy rested on mathematical and conceptual similarity, but was not derived from any single unifying principle. In the meantime, the Bayes--Laplace rule has been shown to be a special case of the constrained maximisation of the Kullback--Leibler relative entropy \cite{Williams:1980,Warmuth:2005,Caticha:Giffin:2006,Douven:Romeijn:2012}. Our result provides the missing piece of the puzzle. Both the Bayes--Laplace and von Neumann--L\"{u}ders rules are special cases of a single epistemic principle of inductive inference (or, in other words, information theoretic state updating). This issue will be discussed in more detail in Section \ref{found.view.label}.

After finishing this paper, we were informed about reference \cite{MPSVW:2010}, where it is shown that a state $\sigma=\sum_iP_i\rho_iP_i$, where $P_i$ are rank $1$ projectors, minimises the functional $D(\rho,\sigma)$. This is a special case of our result for the weak collapse rule. The generalisation to arbitrary projectors is suggested in \cite{Coles:2012}, but without a proof or an indication of a method of proving this statement. The technique used by us to prove a general theorem is essentially different from one applied in \cite{MPSVW:2010} (and it shows that this result for the rank $>1$ case is more substantial and nontrivial than for the rank~1 case).

A closely related paper \cite{Kostecki:2014:lqj} deals with the same type of problem as addressed here, but using a different mathematical approach, allowing for treatment of the infinite dimensional case. Further conceptual and mathematical discussion associated with the results of both papers is carried out there and in \cite{Kostecki:2015:beyond}. A recent work \cite{MunkNielsen:2015} proves that a partial trace is also a constrained maximiser of quantum relative entropy.

\section{The setup}

We will consider the finite dimensional case. Hence, quantum states will be identified with non-negative matrices of trace $1$, which form the convex set $\Dens$ in the space of all hermitian $n\times n$ complex matrices.

The function $\Dkl(\cdot,\cdot)$  is jointly convex in both arguments \cite{Lindblad:1974}, which implies that $\Dkl(\rho,\cdot)$ is convex on $\Dens$ for all $\rho\in\Dens$. Due to the finite dimensionality of the problem, we can use the first order condition for the existence of a minimum of a convex function (see e.g. \cite{Borwein:Vanderwerff:2010}, Theorems 1.2.7 and 2.2.1): if $\Vari$ is a convex subset of a finite dimensional topological vector space, and $f:\Vari\ra\RR$ is convex
then $x$ is a global minimum of $f$ on $\Vari$ if and only if all directional derivatives of $f$ at $x$ are nonnegative.

For a function differentiable at $x$ this condition states that if $x$ is in the interior of $\Vari$ then the derivatives of $f$ need to vanish. If $x$ belongs to some strata of the boundary of $\Vari$ then all tangential derivatives need to vanish whereas derivatives in inward transversal direction need to be nonnegative.


In our minimisation problem we have a subspace $\Vari\subset\Dens$ of density matrices that is defined by a linear equation and thus is a subsimplex. The function $\Dkl(\rho,\cdot)$ restricts to a convex and differentiable function on $\Vari$ and we want to find its minimum. Thus we simply differentiate in the directions preserving $\Vari$ and set the derivatives to be positive. We will denote this condition by
\begin{equation}
C^\rho_\Vari(\cdot):=\partial_\Vari \Dkl(\rho,\cdot)=\partial_\Vari\tr(\rho\ln(\cdot))\geq 0.
\end{equation}
The next two sections will be concerned with evaluating this set of equations.
\section{Weak collapse}
In the case of a weak collapse due to the measurement of $\x=\sum_i\lambda_iP_i$, the constraint set is given by the block diagonal density matrices,
\begin{equation}
\Vari_w:=\{\sigma\in\Dens\mid[P_i,\sigma]= 0\;\;\forall P_i\}.
\label{Vari.w.condition}
\end{equation}
The condition \eqref{Vari.w.condition} is equivalent with $\sigma\in\Vari_w$ if{}f $\sigma = \sum_i P_i \sigma P_i$, as well as with $\sigma\in\Vari_w$ if{}f $[\x,\sigma]=0$ (see \cite{Herbut:1969} for a discussion).

We can parametrise $\Vari_w$ in terms of the singular value decomposition of $\sigma$. Every element of $\Vari_w$ is of the form
\begin{equation}
	\sigma = U\Diag U^*,
\end{equation}
with  $\Diag$ a trace $1$ diagonal matrix with positive entries, $U$ a unitary that is a product $U=\prod_i U_i$, where $U_i$ is an identity on the range of $\II-P_i$. We have that $[U,P_i]=0$, $[\Diag,P_i]=0$, and thus, writing $\sigma_i=\sigma|_{\ran(P_i)}$ we have $f(\sigma)=\bigoplus_if(\sigma_i)=\bigoplus_if(\sigma_i)$; that is, functions (in the sense of the functional calculus) act blockwise on the space $\Vari_w$.


Let us consider first the variation $\partial_{\Vari_w}\tr(\rho\ln(\cdot))=0$ in the direction parametrised by the $U_i$. Given a function on a Lie group $f(U)$ we can take the directional derivative by looking at the parameter derivative of a one parameter group of diffeomorphisms on $U$. As multiplication in a Lie group is differentiable we can pick the one parameter group of diffeomorphisms generated by left multiplication with the one dimensional subgroup $\exp(tL)$,
\begin{equation}
	\phi_t(U) = \exp(tL)(U).
\end{equation}
We then define the directional derivative in direction $L$ as the derivative of the pushforward of $f$ along $\phi_t$, $\partial_Lf(\cdot)=\frac{\dd}{\dd t}\phi_t^\sharp f(\cdot)|_{t=0}$. For a function that is the trace of $U$ in a particular representation this can be easily evaluated:
\begin{eqnarray}
&\frac{\dd}{\dd t}\phi_t^\sharp\tr(AU)|_{t=0}=\frac{\dd}{\dd t}\tr(A\ee^{tL}U)|_{t=0}=\nonumber\\
&\sum_{i,j}(A)_{ij}\frac{\dd}{\dd t}(\ee^{tL}U)_{ji}|_{t=0}=\sum_{i,j}(A)_{ij}(LU)_{ji}=\nonumber\\
&\tr(ALU).
\end{eqnarray}
A straightforward calculation shows that we further have $\frac{\dd}{\dd t}\phi_t^\sharp\tr(AUBU^*)|_{t=0} = \tr(ALUBU^*)-\tr(AUBU^*L)$.

Note that $[L_i,P_j]=0$, and in particular $L_iP_j=\delta_{ij}L_i$. The derivative then takes the form
\begin{align}
\partial_{L_i} \sum_j \tr(P_j \rho P_j U_j \ln \Diag _j U_j^*) &=\nonumber\\
\tr(\rho_i L_i \ln \sigma_i) - \tr(\rho_i \ln (\sigma_i) L_i) &= 0,
\end{align}
\begin{equation} 
\tr(L_i [\ln \sigma_i,\rho_i])=0.
\end{equation}

We thus see that if $\sigma$ and $\sum_i P_i \rho P_i$ are concurrently diagonalisable, the above equation vanishes. In fact, since $[\ln\sigma_i,\rho_i]$ is traceless and $\{L_i,\ii L_i\}$ spans the space of all traceless matrices in the $i$-th matrix block, this is also a necessary condition.

Let us next consider the variation in the direction of the spectrum, that is the direction of $\Diag $. We are interested in the case where $\sigma$ and $\sum_iP_i\rho P_i$ are concurrently diagonalisable. Let $\kappa^{\sigma}_k$ and $\kappa^{\rho}_k$ be the eigenvalues of $\sigma$ and $\sum_i P_i \rho P_i$ respectively.  If $\kappa^\rho_i\not=0$ and $\kappa^\sigma_i =0$ then $\Dkl(\rho,\sigma) = \infty$, so this can not be the minimum if a state with finite relative entropy exists, and we can disregard this case here.

Let us first consider the case that all $\kappa^\rho_i\not=0$.
We have the condition
\begin{equation}
	\partial_{\Diag _\sigma}\Dkl(\rho,\sigma)=\partial_{\Diag _\sigma} (-\tr(\Diag _\rho \ln \Diag _\sigma)) = \partial_{\Diag _\sigma} \sum_k -\kappa^{\rho}_k \ln \kappa^{\sigma}_k = 0.
\end{equation}
The derivatives $\partial_{\Diag _\sigma}$ have to preserve the trace. An overcomplete basis of such derivatives is given by $\partial_{\kappa^\sigma_k}-\partial_{\kappa^\sigma_l}$. Thus, for all $k,l,m$
\[
\left(\partial_{\kappa^\sigma_k} - \partial_{\kappa^\sigma_l}\right)\sum_m -\kappa^{\rho}_m \ln \kappa^{\sigma}_m = 0,
\]
\begin{equation}
\frac {\kappa^{\rho}_l}{\kappa^\sigma_l} - \frac {\kappa^{\rho}_k}{\kappa^\sigma_k} = 0, 
\end{equation}
So, the ratios of the eigenvalues of $\sum_iP_i\rho P_i$ and $\sigma$ are fixed. As they both are trace $1$, this implies they are the same.
 
Let assume now that $I$ is the index set of all $i$ such that $\kappa^\rho_i=0$.
If this set is nonempty then the above conditions cannot be satisfied. 
However, there is still a possibility that the minimum is on the boundary. The condition for the minimum on the boundary is weaker than the above. Namely, all derivatives in directions pointing in toward the set need to be positive. Such directions can be written as a linear combination 
\begin{equation}
 \sum_{i\in I, j\notin I}\alpha_{ij}(\partial_{\kappa^\sigma_i}-\partial_{\kappa^\sigma_l})
 +\sum_{i,j\notin I}\alpha_{ij}(\partial_{\kappa^\sigma_i}-\partial_{\kappa^\sigma_l})
\end{equation}
with $\alpha_{ij}\geq 0$ for $i\in I$, $j\notin I$ and otherwise $\alpha_{ij}$ arbitrary, since the derivatives with negative
coefficients at $\partial_{\kappa^\sigma_i}$ would otherwise point outside the set. For that it is enough to check basis derivatives
\begin{align}
\forall_{i\in I,j\notin I}\left(\partial_{\kappa^\sigma_i} - \partial_{\kappa^\sigma_j}\right)\sum_m -\kappa^{\rho}_m \ln \kappa^{\sigma}_m &\geq 0\nonumber\\
\Rightarrow \forall_{l\notin I}\frac{\kappa^{\rho}_l}{\kappa^\sigma_l}&\geq 0,\\
\forall_{i,j\notin I}\left(\partial_{\kappa^\sigma_i} - \partial_{\kappa^\sigma_j}\right)\sum_m -\kappa^{\rho}_m \ln \kappa^{\sigma}_m&=0\nonumber\\
\Rightarrow \forall_{k,l\notin I}
\frac {\kappa^{\rho}_l}{\kappa^\sigma_l} - \frac {\kappa^{\rho}_k}{\kappa^\sigma_k} &= 0, \label{bulk.minimum.eq}
\end{align}
and we see this is a global minimum. 

Recall that if $\kappa^\rho_i \neq 0$ when $\kappa^\sigma_i = 0$ then $\Dkl(\rho,\sigma)=\infty$. We now also need to consider the case that
$\kappa^\rho_i = 0$ when $\kappa^\sigma_i \not= 0$.
In that case we would get the full derivatives in the $i$ direction, thus the equations \eqref{bulk.minimum.eq} apply, which can not be satisfied unless all $\kappa^\sigma_j = 0$, which can not occur in $\Dens$.

Combining this with the above we have that
\begin{equation}
C_{\Vari_w}^\rho(\sum_iP_i\rho P_i)=\partial_{\Vari_w}\Dkl(\rho,\sum_iP_i\rho P_i) \geq 0.
\label{zeroset.condition.Vari.w}
\end{equation}
The state $\sigma=\sum_iP_i\rho P_i$ is the only state $\sigma$ satisfying $C^\rho_{\Vari_w}(\sigma) \geq 0$. The set $\Vari_w$ is convex, so from \eqref{zeroset.condition.Vari.w} and convexity of $\Dkl(\rho,\cdot)$, this is the unique global minimum.

\section{Strong collapse}

The conditions defining `strong' collapse that were specified in Introduction lead us to a troubling situation, because for such states (containing zero eigenvalues) the relative entropy is almost always infinite. We will overcome the problem by deriving a generalised version of the strong collapse rule that is a quantum counterpart of Jeffrey's rule. The ordinary strong collapse rule will be then obtained by a limiting procedure.

Consider a constraint set given in terms of $p_i\in\RR$ such that $\sum_ip_i=1$ by
\begin{equation}
\Vari_s:=\{\sigma\in\Dens\mid[P_i,\sigma]=0,\;\;\tr(\sigma P_i)=p_i\;\forall P_i\},
\label{Vari.s.condition}
\end{equation}
where $\{P_i\mid i\in I\}$ is again determined by the spectral decomposition of an observable $\x=\sum_{i\in I}\lambda_iP_i$. The set \eqref{Vari.s.condition} can be interpreted as encoding the knowledge that the measurement outcome $\lambda_i$ corresponding to a projection $P_i$ occurs with a probability $p_i$.

Here we encounter a problem. If we have a $p_i$ nonzero but $\tr(\rho P_i)=0$, then every state in $\Vari_s$ will have relative entropy $-\infty$ to $\rho$. Moreover, even if we subtract the infinite constant, we find that the regularised distance does not depend on the state in the block $P_i$ and there is no unique minimum. We thus will always assume that $\tr(\rho P_i)\neq0$ for $p_i \neq 0$.

The variation in the $U_i$ direction goes through as before. However the variation in the direction of the spectrum changes in that a basis is now given in terms of $\partial_{\kappa^{\sigma_i}_k} - \partial_{\kappa^{\sigma_i}_l}$, with $\kappa^{\sigma_i}_k$ and $\kappa^{\sigma_i}_l$ belonging to the same block $P_i$ and thus being eigenvalues of $\sigma_i$. Thus only the fractions of eigenvalues within each block are fixed. This implies that the eigenvalues of $\sigma_i$ are uniformly scaled relative to the eigenvalues of $\rho_i$. The condition $\sum_k\kappa^{\sigma_i}_k=p_i$ fixes $\sigma_i$ to be $p_i\rho_i/\tr(\rho_i)$.

This shows that
\begin{equation}\label{eq:vari-Jeffrey}
C^\rho_{\Vari_s}\left(\sum_ip_i\frac{P_i\rho P_i}{\tr(P_i\rho P_i)}\right) \geq 0.
\end{equation}
The state $\sigma=\sum_ip_i\frac{P_i\rho P_i}{\tr(P_i\rho P_i)}$ is the only state $\sigma$ satisfying $\partial_{\Vari_s}\Dkl(\rho,\sigma) \geq 0$.

The strong collapse is a limiting case of the above projection, with all $p_i$ going to zero except of one, $p_1$, corresponding to a projection $P_1$ that, in turn, corresponds to a measurement result given by an eigenvalue $\lambda_1$. We obtain this by taking the weak continuous limit.
\begin{eqnarray}
	\lim_{p_2,p_3,\ldots\ra0}\arginff{\sigma\in\Vari_s}{\Dkl(\rho,\sigma)}&=
	\lim_{p_2,p_3,\ldots\ra0}\sum_ip_i\frac{P_i\rho P_i}{\tr(P_i\rho P_i)}\nonumber\\&=
	\frac{P_1\rho P_1}{\tr(P_1\rho P_1)}.
\end{eqnarray}
Note that in the finite dimensional case that we consider here the weak topology and norm topology coincide.

\section{The foundational view\label{found.view.label}}

In the orthodox formulation of quantum mechanics the `collapse rules' are \textit{postulated}. Thus, they are not deduced from any other more fundamental principle. They can be derived from several different conditions, see \cite{Busch:Lahti:Mittelstaedt:1991,Busch:Grabowski:Lahti:1995:FP} for a review, but none of these conditions possesses the status of a fundamental principle of quantum theory. The weak collapse rule can be derived by taking the tensor product with an auxilliary state, followed by unitary evolution and a partial trace. This may serve as a derivation independent of interpretational issues (when this procedure is interpreted as an interaction with some ontic environment, it is usually considered as an instance of decoherence). However, no such construction exists for the strong rule. This fact, as well as the unclear relationship between the strong collapse rule and unitary evolution, renders the orthodox mathematical foundations conceptually insufficient, asking for further insights.

In general, an ontic interpretation of the quantum state leads to considering quantum collapse as a change of the ``state of being'' of some ``material object/thing''. On the other hand an epistemic interpretation leads to considering quantum collapse as a change of the ``state of information'' of some ``experiencing user/agent''. (There also is a corresponding difference in the meaning of the term `measurement'.) In particular, the dynamical reduction approach of \cite{Ghirardi:Rimini:Weber:1986}, belongs to the former class, providing an ontic \textit{explanation} by means of a general dynamical principle from which the quantum collapse rule is derived. On the other hand, an epistemic \textit{interpretation} of collapse rules as quantum mechanical analogues of the Bayes--Laplace rule $p(x)\mapsto p(x)p(b|x)/p(b)$ was proposed in \cite{Bub:1977,Bub:1979:measurement,Fuchs:2002,Jacobs:2002}. However, no epistemic \textit{explanation}, understood as a derivation from some fundamental principle of information theory (or statistical inference theory) has been offered. Our paper (as well as the closely related paper \cite{Kostecki:2014:lqj}) provides such a derivation.

Following the postulates of \cite{Kostecki:2010:AIP,Kostecki:2015:beyond} (which aim at reapproaching the foundations of quantum theory in the spirit of \cite{Jaynes:1957,Kullback:1959,Jaynes:1979:where:do:we:stand,Shore:Johnson:1980}), we demonstrated that the mapping to the unique solution of constrained minimisation of the relative information $\Dkl$,
\begin{equation}
	\rho\mapsto\arginff{\sigma\in\Q}{\Dkl(\rho,\sigma)},
\label{MRE.eq}
\end{equation}
can serve as the general principle of quantum state change due to the acquisition of new information (represented by the constraints $\Q$). This amounts to selecting the quantum state that is the least distinguishable from the original state among all states that are in a strict agreement with the new knowledge (represented by the constraints). 

In order to derive the quantum collapse rules from the principle \eqref{MRE.eq}, we needed to identify the information theoretic constraints that \textit{define} the situations of weak and strong collapse. The `weak' collapse amounts to encoding the information that a specific observable $\x$ has been subjected to measurement. A quantum state $\sigma$ that carries such information has to be compatible with the possibility of measuring all eigenvalues of $\x$ precisely. Such a situation can be characterised by the condition $[\sigma,\x]=0$ (or, equivalently, $[P_i,\sigma]=0$ $\forall P_i$). The `strong' collapse should additionally result in a state that would reproduce the result of measurement of a particular eigenvalue with certainty (that is, with probability equal $1$). That is, given a projector $P$ encoding the outcome $\lambda$ of the measurement, the post-collapse density operator $\sigma$ should satisfy the condition of a `weak' collapse, as well as $\tr(P\sigma)=1$. This provides an interesting general insight into a structure of quantum theory: why it is possible to use \eqref{MRE.eq} in order to derive various quantum state change rules without assuming the probabilistic interpretation carried under the label of the `Born rule', the latter seems to be required for justification of the choice of constraints leading to a specific class of rules, including `strong' collapse.

Our results can be considered as a quantum counterpart of derivations \cite{Williams:1980,Warmuth:2005,Caticha:Giffin:2006,Douven:Romeijn:2012} of the Bayes--Laplace rule from the constrained maximisation of the Kullback--Leibler relative entropy \cite{Kullback:Leibler:1951}, $\Sre(p,q):=-\int_\X\mu(x)p(x)\log(p(x)/q(x))$, where $x\in\X$, while $p$ and $q$ are densities of probability measures with respect to a measure $\mu$ on $\X$. The functional $\Sre(p,q)$ is a special case of Umegaki's quantum relative entropy $\Sre(\sigma,\rho)$ for discrete $\X$ and $[\sigma,\rho]=0$. This strengthens the analogy between the Bayes--Laplace and the von Neumann--L\"{u}ders rules: they are just two special cases of a single general principle of inductive inference, given by \eqref{MRE.eq}. From the Bayesian perspective, the state $\rho$ is a prior, while $\sigma$, satisfying the constraints and maximising $\Sre(\rho,\sigma)$, is a posterior.

\section{Remarks}

It has been known for quite a long time (see e.g. \cite{Wehrl:1978}) that a `weak' collapse \textit{leads to} an increase of the \textit{absolute} entropy $-\tr(\rho\log\rho)$. Our result uncovers an unexpectedly strong inverse of this fact: a `weak' collapse \textit{is a result of} maximisation of the \textit{relative} entropy $-\tr(\rho\log\rho-\rho\log\sigma)$ under specific constraints.

All earlier results on derivation of weak and strong collapse rules from minimisation of two point functionals on the space of quantum states \cite{Herbut:1969,Marchand:Wyss:1977,Benoist:Marchand:Yourgrau:1977,Gudder:Marchand:Wyss:1979,Hadjisavvas:1981,Dieks:Veltkamp:1983,Marchand:1983,Marchand:1983:Milano,Raggio:1984} were obtained for (various) \textit{symmetric} quantum information distances. The importance of our result stems from the importance of (the negative of) Umegaki's relative entropy in quantum information theory as opposed to symmetric quantum information distances, which do not carry a similar semantic significance. This statement can be approached either axiomatically or pragmatically. On the axiomatic side, $D(\rho,\sigma)$ is characterised \cite{Petz:1992:characterisation} by the direct sum property, invariance under automorphisms (so, in particular, unitaries), additive decomposition under conditional expectations (onto subalgebra), and measurability over the state space. These properties eliminate all above symmetric information distances. An extensive discussion of the reasons for (and, in particular, applications of) these properties can be found in \cite{Ohya:Petz:1993}. Furthermore, $D(\rho,\sigma)$ is a direct quantum generalisation of $-\Sre(p,q)$, which can be also characterised `dynamically' as a unique functional $\Phi(p,q)$ such that the mapping $q\mapsto\arginff{p\in C}{\Phi(p,q)}$ satisfies a few very reasonable desiderata for information processing \cite{Shore:Johnson:1980,Csiszar:1991,Caticha:Giffin:2006}. On the pragmatic side, $D(\rho,\sigma)$ is widely used in quantum information theory as the most fundamental measure of distinguishability of quantum states (see e.g. \cite{VPRK:1997,Nielsen:Chuang:2000,Schumacher:Westmoreland:2002,Vedral:2002,Keyl:2002}). Hence, from the perspective of quantum information theoretic approaches to foundations of quantum theory, our results provide an essential, new perspective on the mathematical form of collapse rules due to quantum measurement.

As noted by one of the referees, this leads to a question whether the results of this paper can be reproduced (or extended) in the setting of generalised probabilistic theories \cite{Barrett:2007,Barnum:Wilce:2009:I}. This setting lacks a general analogue of the collapse rules, but it allows us to introduce a well defined notion of information distance \cite{Short:Wehner:2010,BBCLSSWW:2010} (which reduces to the Umegaki and Kullback--Leibler distances in quantum mechanical and probabilistic case, respectively).\footnote{This problem seems to reflect quite similar issue in quantum logic, where the collapse rules are also not a part of the basic framework, so they require an additional justification or derivation (see e.g. \cite{Stairs:1982}), pointing towards a possibility of some more fundamental principle. Such derivations were provided (under some assumptions) by means of minimisation of suitable symmetric distances (see e.g. \cite{Hadjisavvas:1981,Dieks:Veltkamp:1983,Raggio:1984}), corresponding (in some cases) to symmetric transition probability functionals. It was shown in \cite{Guz:1980} that nonsymmetric transition probability \cite{Mielnik:1969} plays more fundamental role in quantum logic, but its relationship to nonsymmetric distances and to the problem of derivation of collapse rules was not investigated.} Hence, the possible extension of our result to generalised probabilistic theories can bring in new foundational insights (in particular---as suggested by a referee---one can ask whether defining a post-measurement state as a minimiser of a specific information distance given some type of constraints preselects some type of theories). The main open technical problem is how to replace the use of block diagonal decomposition and variational analysis of the spectrum of operators by some other method. It may be possible that a restriction to a subclass of theories satisfying some sort of spectral condition (see e.g. \cite{Alfsen:Shultz:2003,Krumm:2015}) will be necessary for this. We hope to return to this problem in another paper.

\section*{Acknowledgments}

We would like to thank Carlos S. Guedes for many important and insightful discussions throughout the development of this result. We thank also Patrick Coles for informing us about \cite{MPSVW:2010,Coles:2012}, and Daniel Ranard for some suggestions and comments. This research was supported in part by Perimeter Institute for Theoretical Physics. Research at Perimeter Institute is supported by the Government of Canada through Industry Canada and by the Province of Ontario through the Ministry of Research and Innovation. This research was also partially financed by the National Science Center of the Republic of Poland (Narodowe Centrum Nauki) through the grant number DEC2011/01/N/HS3/03273.


\end{document}